\documentclass[11pt]{article}
\usepackage{epsfig,psfig,float,amssymb,latexsym}
\usepackage[notref,notcite]{}
\newcommand{\NP}[1]{ Nucl.\ Phys.\ {#1}}

\newcommand{\PL}[1]{ Phys.\ Lett.\ {#1}}

\newcommand{\PR}[1]{Phys.\ Rev.\ {#1}}
\newcommand{\PRL}[1]{ Phys.\ Rev.\ Lett.\ {#1}}

\newcommand{\vs}{\vspace{-0.2cm}}

\newcommand{\La}{{\cal L}}

\newcommand{\Ima}{{\rm Im}\,}
\newcommand{\Rea}{{\rm Re}\,}

\newcommand{\be}{\begin{equation}}
\newcommand{\ee}{\end{equation}}
\newcommand{\ba}{\begin{eqnarray}}
\newcommand{\ea}{\end{eqnarray}}

\newcommand{\nn}{\nonumber}

\begin{document}
\hfill{IFIC$-$01$-$0301}

\hfill{FTUV$-$01$-$0301}
\begin{center}
{\Huge{\bf{Chiral unitary theory: \\
\vspace{0.4cm}
application to nuclear problems}}}
\end{center}
\vspace{.5cm}

\begin{center}
{\huge{E. Oset$^1$, D. Cabrera$^1$, H.C. Chiang$^2$, C. Garcia Recio$^3$,
 S. Hirenzaki$^4$, S.S. Kamalov$^5$,
 J. Nieves$^3$, Y. Okumura$^4$, A. Ramos$^6$, 
H. Toki$^7$ and M.J. Vicente Vacas$^1$}}
\end{center}

\begin{center}
{\small\it{$^1$Departamento de F\'{\i}sica Te\'orica and IFIC,
Centro Mixto Universidad de Valencia-CSIC,\\
Institutos de Investigaci\'on de Paterna, Apdo. correos 2085,\\
46071, Valencia, Spain. \\
$^2$Institute of High Energy Physics, Academia Sinica, Beijing, China.\\
$^3$Departamento de Fisica Moderna, Universidad de Granada, Spain.\\
$^4$Physics Department, Nara Women University, Nara, Japan.\\
$^5$Laboratory of Theoretical Physics, JINR Dubna, Russia.\\
$^6$Departament d'Estructura i Constituents de la Materia, Universitat
de Barcelona, Spain\\
$^7$RCNP, Osaka University, Osaka, Japan}}
\end{center}

\vspace{1cm}

\begin{abstract}
{\small{ In this talk we briefly describe some basic elements of chiral
perturbation theory, $\chi PT$, and how the implementation of unitarity and
other novel elements lead to a better expansion of the $T$ matrix for meson
meson and meson baryon interactions.  Applications are then done to the $
\pi \pi $ interaction in nuclear matter in the scalar and vector channels,
antikaons in nuclei and $K^-$ atoms, and how the $\phi$ meson properties are 
changed in a nuclear medium.
}}
\end{abstract}

\section{Introduction}

  Nowadays it is commonly accepted that QCD is the theory of the strong
interactions, with the quarks as building blocks for baryons and mesons, and
the gluons as the mediators of the interaction. However, at low energies typical
of the nuclear phenomena, perturbative calculations with the QCD Lagrangian are
not possible and one has to resort to other techniques to use the information of
the QCD Lagrangian. One of the most fruitful approaches has been the use of
chiral perturbation theory, $\chi PT$. The theory introduces effective
Lagrangians which involve only observable particles, mesons and baryons,
respects the basic symmetries of the original QCD Lagrangian, particularly
chiral symmetry, and organizes these effective Lagrangians according to the
number of derivatives of the meson and baryon fields.

The lowest
order chiral Lagrangian for the meson meson interaction, invariant under 
Lorentz transformations, parity and 
charge conjugation with only two derivatives and linear in the quark masses 
is \cite{xpt}

\begin{equation}
\label{L2}
\La_2=\frac{f^2}{4}< D_\mu U^\dagger \, D^\mu U+U^\dagger \,\mathcal{M}+
\mathcal{M}^\dagger \,U> \ ,
\end{equation}
where $<>$ means $SU(3)$-flavour trace, with
$U(\Phi)=\exp(\frac{i\sqrt 2 \Phi}{f})$, $D^\mu U$ the covariant derivative of
$U$ (normal derivative in the absence of external fields), and 

\begin{equation}
\label{Fi}
\Phi=\frac{\vec{\lambda}}{\sqrt{2}}\, \vec{\phi}=\left(
\matrix{
\frac{1}{\sqrt{2}} \pi^0+\frac{1}{\sqrt{6}}\eta_8 & \pi^+ &
K^+ \cr
\pi^- & -\frac{1}{\sqrt{2}}\pi^0+\frac{1}{\sqrt{6}}\eta_8 & K^0 \cr
K^- & \bar{K}^0 & -\frac{2}{\sqrt{6}}\eta_8}
\right)
\end{equation}

The mass matrix $\mathcal{M}$ is given in terms of the meson masses and in the
limit of equal up and down quark masses reads

\be
\label{mmatrix}
\left(
\matrix{
m_\pi^2 & 0 & 0\cr
0& m_\pi^2 & 0 \cr
0&0& 2 m_K^2 - m_\pi^2} 
\right) \ .
\ee

The meaning of the constant $f$ can be appreciated when calculating from
the
lowest order Lagrangian the axial current. Then $f$
becomes the pion decay constant in the chiral limit, which is about 93 MeV.

The next to leading order Lagrangian, $\La_4$, is constructed with the same building 
blocks as $\La_2$, preserving Lorentz invariance, parity and charge 
conjugation and explicit formulae can be seen in \cite{xpt}. They contain four
derivatives on the meson fields or meson masses to the fourth power and for the
purpose of meson meson interaction contain eight $L_i$ free parameters which 
are adjusted to the data or alternatively derived in some models. 

  Chiral perturbation theory up to fourth order (in the number of derivatives) 
 consists in the perturbative field theoretical calculation from the lowest 
 order Lagrangian, which involves loop diagrams and divergences. These are 
 cured by the introduction of the fourth order Lagrangian which also leaves 
 some residual finite contribution to the amplitudes. Review papers on this 
 issue can be seen in \cite{pich,ulf,ecker}.

The inclusion of baryons in the chiral formalism is done in a similar way than
the one used for the mesons. We consider here the octet of
baryons

\begin{equation}
B=\left(
\matrix{\frac{1}{\sqrt{2}}\Sigma^0 +\frac{1}{\sqrt{6}}\Lambda^0
& \Sigma^+
& p \cr \Sigma^- &
-\frac{1}{\sqrt{2}}\Sigma^0 +\frac{1}{\sqrt{6}}\Lambda^0 & 
n \cr \Xi^- & \Xi^0 & -\frac{2}{\sqrt{6}}\Lambda^0}
\right)
\end{equation} 

The lowest order baryon-meson Lagrangian with at most two baryons can
be written as:

\begin{eqnarray}
\label{BaryonL}
\La_1=<\bar{B}i\gamma^\mu \bigtriangledown_\mu B >-M_B
<\bar{B}B>&+&\frac{1}{2}D<\bar{B}\gamma^\mu \gamma_5
\left\{u_\mu,B \right\}>\\
&+&\frac{1}{2}F<\bar{B}\gamma^\mu \gamma_5 [u_\mu,B] >\nonumber
\end{eqnarray}
where
\begin{equation}
\label{CovDer}
\bigtriangledown_\mu B=\partial_\mu B+[\Gamma_\mu,B] \ ,
\end{equation}
with $\Gamma_\mu$  defined below and 
$D+F=g_A=1.257$ and $D-F=0.33$ \cite{pich}. We have
\begin{equation}
\label{Gammamu}
\Gamma_\mu=\frac{1}{2}\left\{ u^\dagger\partial_\mu u+
u\partial_\mu u^\dagger \right\}
\ee
with $u$ such that $u^2=U$ and 
\ba
u_\mu&=&i u^\dagger D_\mu U u^\dagger=u^\dagger_\mu  
\ea

\section{Chiral unitary theory}

   The perturbative series in powers of $p^2$ of the meson momenta is slowly
   convergent and in the case of $\pi\pi$ interaction only works up to values of
   the centre of mass energy of the two meson system of the order of 400 MeV.
   This range is such that one misses the meson resonances like the $f_0(980)$,
   $a_0(980)$, $\sigma(500)$, $\rho(770)$, $K^*(900)$, etc. On the other hand,
   the intrinsic nature of the perturbative scheme makes it impossible to obtain
   the infinite value for the $T$ matrix which characterizes the appearence of a
   resonance. Nonperturbative schemes are hence necessary to obtain these
   singularities. Several methods have been proposed recently in the context of
   the chiral Lagrangians \cite{ramonet,iam,juan,nsd}. Although they are
  technically different, they share many things in common and the essence of the
   method can be qualitatively understood by looking at how the inverse 
   amplitude method in coupled channels works in \cite{iam}.
  Unitarity in coupled channels reads in our normalization:
\begin{equation}
\Ima T_{if} = -T_{in} \, \rho_{nn} \, T^*_{nf}
\label{Tunit}
\end{equation} 
where $\rho$ is a real diagonal matrix whose elements account 
for the phase space of the two meson intermediate states $n$ which are
physically accessible. With our normalization, $\rho$ 
is given by

\begin{equation}
\label{1.0}
\rho_{nn}(s)=\frac{k_n}{8 \pi \sqrt{s}} \theta (s-(m_{1n}+m_{2n})^2)
\end{equation}
where $k_n$ is the on shell center mass (CM) momentum of the meson in the 
intermediate state $n$ and $m_{1 n}, m_{2 n}$ are the masses of the two mesons 
in this state.

Isolating $\rho$ from eq. (\ref{Tunit}) one has:

\begin{eqnarray}
\rho &=& -T^{- 1} \cdot \Ima T \cdot T^{* - 1}\nn \\
&=& -\frac{1}{2 i} T^{- 1}\cdot (T - T^*)\cdot T^{* - 1} \nn\\
&=& -\frac{1}{2 i} (T^{- 1 *} - T^{- 1}) = \Ima T^{- 1}
\label{ImG}
\end{eqnarray}

 From eq. (\ref{ImG}) we can write:

\begin{equation}
\label{1.1}
T^{-1}=\Rea T^{-1}+i \rho
\end{equation}

Once we are at this point we realize that unitarity gives $Im T^{-1}$ for free
independently of the dynamics. Then it also looks most natural to expand 
$T^{-1}$, instead of $T$, in powers of
$p^2$ . Indeed, whenever $T$ has a pole $T^{-1}$ will have a zero,
but an expansion in a power series around zero is not a problem while we can 
not do
it around a singularity. These are hence the basic elements introduced in these
nonperturbative methods:
1) Unitarity is satisfied exactly,
2) Coupled channels are used,
3) The chiral expansion is made in the inverse of the $T$ matrix instead of $T$.

  These ingredients are sufficient to allow one to reproduce all basic 
  features of the meson
 meson data up to 1.2 GeV, using the same Lagrangians, up to order four, used in
  chiral perturbation theory, as shown recently in \cite{phipipi}. The method
  used in \cite{ramonet} is identical, except that only one channel was used.
  In that method one can obtain the $\sigma(500)$ in the L=I=0 sector but not
  the $f_0(980)$ and $a_0(980)$ which require the explicit inclusion of the
  $\bar{K}K$ channel.  In \cite{juan} the Bethe Salpeter equation is used to
  unitarize in a single channel and provides a justification of the method
  followed 
  in \cite{npa}, where using the Bethe Salpeter equation and the on shell
  amplitudes at order $O(p^2)$  a good agreement with the data in the
  scalar sector could be obtained by means of only one regularizing parameter.
  In the work \cite{nsd} a different idea is followed since the input is now
  based on the lowest order chiral Lagrangians plus the explicit fields
  (essentially vector mesons) which would correspond to preexisting states
  ($\bar{q}q$) of QCD. These are states which would remain in  the absence of 
  the meson meson scattering
  generated by the residual interaction of the lowest order chiral
  Lagrangian. This follows the idea of \cite{rafael} that the fourth order
  chiral Lagrangian is nothing but the reflection of the exchange between the
  mesons of these preexisting resonances. 
  
    In the $\bar{K}N$ system the use of the chiral Lagrangians in the meson
    baryon sector together with the unitarization in coupled channels also
    proves very efficient and lead to excellent results for all the  $K^-p$
    reactions at low energies using the Bethe Salpeter equation and the lowest order
  Lagrangian. Only a cut off introduced to regularize the loop functions is
  needed and one also obtains  the $\Lambda(1405)$ resonance in L=0 and I=0 as
  a meson baryon resonance, generated  dynamically through the multiple
   scattering
  of the meson baryon states induced by the Bethe Salpeter equation \cite{kaon}.
  Similar findings were first obtained in \cite{kaiserk} using however a more
  restricted space of coupled channels but introducing higher order
  Lagrangians to cope for those missing channels. 
  
  A recent review of these methods with many results  can be 
  seen in \cite{review}.

\section{Application to nuclear problems}
\subsection{The $\pi\pi$ interaction in the nuclear medium in the scalar sector}

The $\pi \pi$ interaction in a nuclear medium in the $L=I=0$
channel ($\sigma$ channel) has stimulated much theoretical work
lately.
It was realized that the attractive P-wave interaction of the
pions
with the nucleus led to a shift of strength of the $\pi \pi$
system
to low energies and eventually produced a bound state of the two
pions around $2 m_\pi  -  10$  MeV \cite{Schuck}. This state
would
behave like a $\pi \pi$
Cooper pair in the medium, with repercussions in several
observable
magnitudes in nuclear reactions \cite{Schuck}. The possibility
that such
effects could have already been observed in some unexpected
enhancement
in the ($\pi, 2 \pi$) reaction in nuclei \cite{2} was also
noticed there.
More recent experiments where the enhancement is seen in the
$\pi^+ \pi^-$
channel but not in the $\pi^+ \pi^+$ channel \cite{3} have added
more
attraction to that conjecture.

  The advent of the chiral unitary methods has added new interest in the
  subject and has allowed one to focus on the implications  of the chiral
  constraints which had been known to be relevant in this kind of studies
  \cite{Rapp}. In \cite{chiang} the $\pi\pi$ interaction in a nuclear medium was
  studied following the lines of \cite{npa}, renormalizing the pion 
  propagators in the medium and introducing vertex corrections for
  consistency. The diagrams considered are depicted in figs. 1,2,3.  The results
  for the imaginary part of the $\pi\pi$ amplitude in L=I=0 are shown in fig. 4.
    One can appreciate that there is an accumulation of strength in the region
    of small invariant masses of the two pion system, with qualitative results
    similar to those found in \cite{Schuck}, but we do not get poles in that
    energy region.  The accumulation of strength at these small invariant
    masses could raise hopes that the enhancement of strength at small invariant
    masses found in the $(\pi,2\pi)$ reactions in nuclei in \cite{3} could be
    explained. However, according to a recent study \cite{manolo}, the 
    small nuclear densities involved in this reaction,
    which is rather peripheral, make the changes found in \cite{chiang}
    insufficient to explain the experimental data.
    
      The work of \cite{chiang} includes only the $\pi\pi$ channel, since one is
 only concerned about the low energy region. This work has been generalized 
 to coupled channels in \cite{manolonew} in order to  make predictions for 
 the modification of the $f_0(980)$ and $a_0(980)$ resonances in a nuclear 
 medium.  One finds there that  both these resonances become wider in the
 medium as the nuclear density increases, with the $a_0(980)$
 eventually melting into a background for densities close to normal nuclear
 matter density. The $f_0(980)$  resonance, which in the free space is
 narrower than the $a_0(980)$, still would keep its identity at these high
 densities but with a width as large as 100 MeV or more.  How to produce these
 resonances in a nucleus in order to check the predictions of these studies is a
 present experimental challenge.

  %   AQUI LAS FIGURAS 2,4,6 DEL TRABAJO DE CHIANG
\begin{figure}[ht!]
\centerline{
\includegraphics[width=0.32\textwidth,angle=-90]{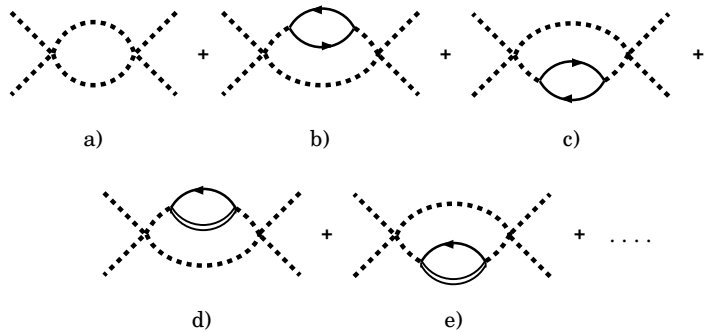}
}
\caption{ Terms appearing in the scattering matrix allowing the
pions to excite
$ph$ and $\Delta h$ components }
\label{fig:pipi1}
\end{figure}   

\begin{figure}[ht!]
\centerline{
\includegraphics[width=0.42\textwidth,angle=-90]{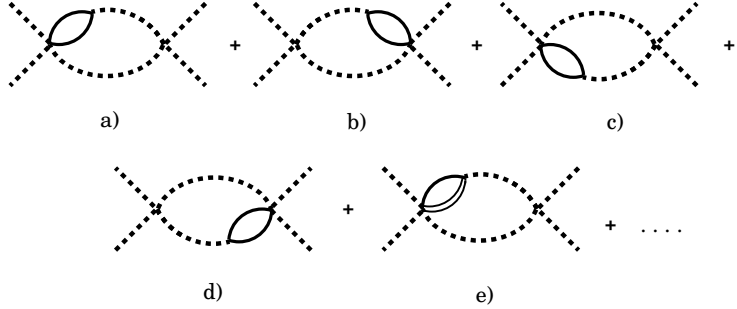}
}
\caption{ Terms of the $\pi \pi$ scattering series in the nuclear
medium related
to three meson baryon contact terms from the Lagrangian of eq.
(5)}
\label{fig:pipi2}
\end{figure} 

\begin{figure}[ht!]
\centerline{
\includegraphics[width=0.25\textwidth,angle=-90]{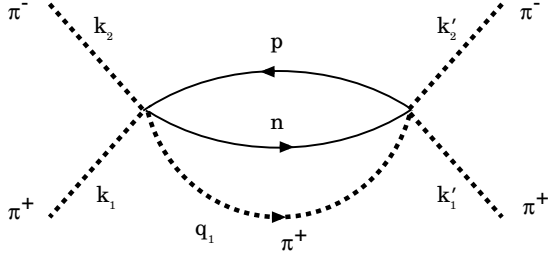}
}
\caption{ Diagram involving the three meson baryon contact terms
of fig. 
\ref{fig:pipi2} 
in each of the vertices }
\label{fig:pipi3}
\end{figure} 

%    AQUI LA FIG. 7 DEL TRABAJO DE CHIANG
 \begin{figure}[ht!]
\vspace*{1cm}
\centerline{
\includegraphics[width=0.55\textwidth,angle=0]{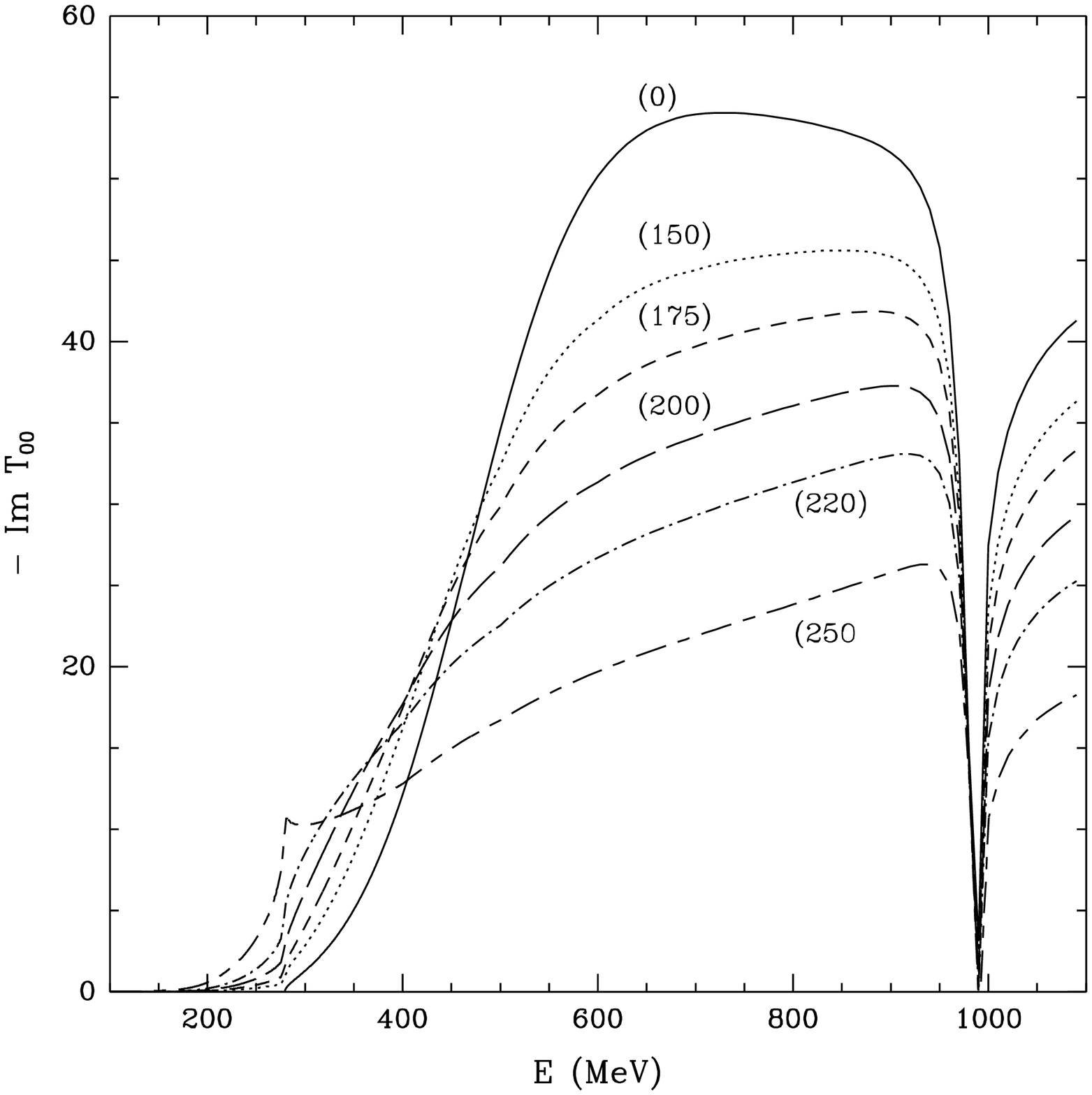}
}
\caption{Im $T_{22}$  for $\pi \pi \rightarrow \pi \pi $
scattering in $J=I=0$
 $(T_{00}$ in the figure) in the
nuclear medium for different values of $k_F$ versus the CM energy
of the 
pion pair. The labels correspond to the values of $k_F$ in MeV. }
\label{fig:pipi4}
\end{figure} 

\subsection{ Isovector $\pi\pi$ scattering and the $\rho$ meson in the nuclear 
medium}
    The modification  of the $\pi\pi$ amplitude in the L=I=1 sector in the
    nuclear medium is equivalent to addressing the modification of the $\rho$
    properties in the medium. Once again this topic has received much
    attention. A good review of the current situation can be
found in ref. \cite{Rapp:1999ej}. Once again we have looked at the problem from
the chiral unitary point of view \cite{Cabrera:2000dx}.  In a first step a combined study of the pion
electromagnetic form factor and the $\pi\pi $ scattering in the vector sector
has been accomplished in \cite{palomar} using the chiral unitary method with
explicit resonances of \cite{nsd}. The nuclear corrections are generated 
similarly
to those discussed above in the scalar sector, introducing both selfenergy
correction in the pion propagators as well as vertex corrections which are
generated by the chiral Lagrangians and are requested by the gauge invariance of
the vector mesons.  
    The results obtained can be summarized in fig. 5, where the real and
 imaginary parts of the $\pi\pi$ amplitude are plotted for normal nuclear matter
 density. The different curves correspond to different choices of a
 regularization parameter, or cut off, which lead to basically the same results
 in free space once the couplings and bare mass of the $\rho$ are changed
 simultaneously. One can see that in the medium the results are also rather
 stable.
 
 \begin{figure}[ht]
\centerline{\includegraphics[width=0.6\textwidth]{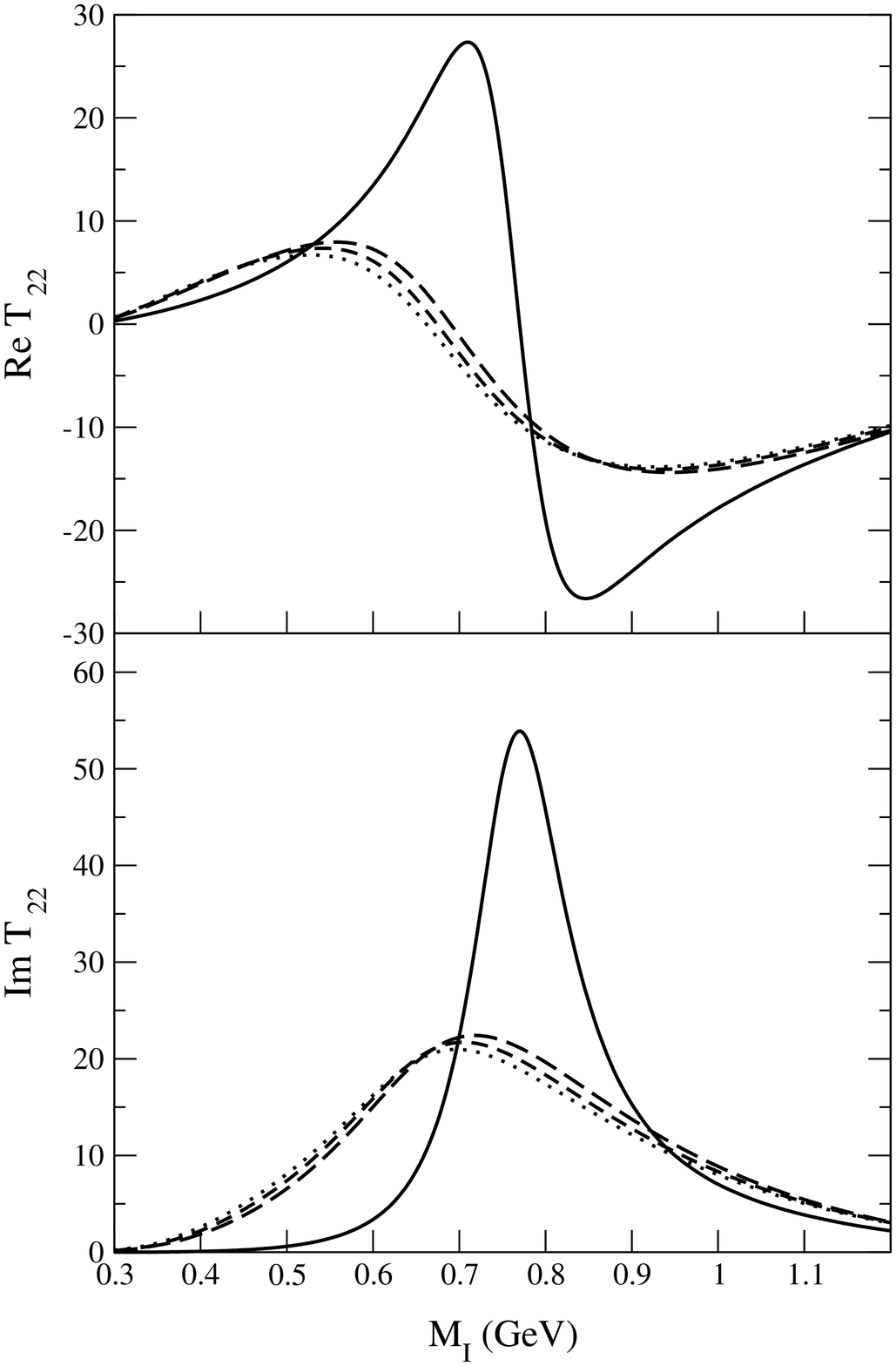}}
\caption{\footnotesize{Real and Imaginary part of the $\pi\pi$ amplitude in
L=I=1 for $\rho=\rho_0$
and several values of $q_{max}$. Short dashed lines stand for $q_{max}=1$ GeV,
long dashed and doted lines being for $q_{max}=0.9,1.1$ GeV respectively.}}
\label{cambio_cut}
\end{figure}

We can observe that the $\rho$ becomes much broader in the nuclear medium but
there is also a change in the position of the peak, implying a shift of the
$\rho$
mass to lower energies by about 50 MeV at $\rho=\rho_0$.  The renormalization of
the $\rho$ still gets contributions from other sources, particularly from the
excitation of the $N^*(1520)h$ by the $\rho$ \cite{Peters:1998va}, since the 
decay of the $N^*(1520)$ resonance into $N\rho$ is one of the important
channels, only reduced in practice because of the limited phase space for the 
decay. The inclusion of this channel produces some extra strength at lower
energies, around 250  MeV below the $\rho$ peak. In \cite{Peters:1998va} the
selfconsistent consideration of the $N^*(1520)$ resonance in the medium, together
with the widening of the $\rho$ width, leads to a broad peak for the spectral
function below the $\rho$ meson peak resulting in a considerable widening of the
$\rho$ strength which would make a shift of 50 MeV irrelevant when compared
to an in medium  width of more than 300 MeV. Present studies at CERN 
\cite{Agakishiev:1995xb,Lenkeit:1999xu} and at lower energies at Bevalac 
\cite{Porter:1997rc} seem to be consistent with a large broadening of the
$\rho$ and more experimental studies are under way at GSI(HADES Collaboration)
  \cite{Friese:1999qm,Bratkovskaya:2000mb}.
  
\section{Kaons in a nuclear medium}
\subsection{$K^-$ deuteron scattering length}
The $\bar{K}N$ interaction is quite strong and when one studies the interaction
of $\bar{K}$ in nuclei important renormalization effects take place. This is
already visible in the interaction of $K^-$ with the deuteron which has been the
subject of much study in the past \cite{landau,koltun,dalitz}.  It was already
known that the evaluation of the $K^-$ deuteron scattering length required the
consideration of multiple scattering of the $K^-$ which was done using Faddeev 
equations. We
have also made some contribution to the field \cite{sabit}, by using again Faddeev
equations in the fixed scatterer approximation which is known to be rather
reliable \cite{deloff}, but using input from the chiral unitary approach of
\cite{kaon} for the elementary amplitudes.  The results are summarized in the 
Table 1, where we show the
results with the impulse approximation (IA), the IA plus double rescattering,
idem plus triple rescattering, the effect of including the charge exchange
($\bar{K^0}$ exchanged between the nucleons), etc.  We can see in the Table that
the full result of the Faddeev calculation is quite different from any of the
approximations, and one can also see that the multiple scattering series does
not converge, which forces the solution by means of the coupled Faddeev 
equations.  We
can also see there that the use of isospin symmetry leads to somewhat inaccurate
results and finally, we also show for comparison the results of \cite{deloff}
which are also quite different than those obtained here, as is also the case for
the results obtained in \cite{gal2}. The main reason lies in the different
elementary amplitudes $\bar{K}N$ provided by the chiral approach with respect to
those used as input in the previous approaches. Experiments to measure this
scattering length are under way in Frascati \cite{guaraldo} and we hope they can
be accomplished in the near future, hence providing further constraints to test
the predictions of these chiral models.

  %\begin{minipage}{5in}
%%%%%%%%%%%%%%%   Table 2 %%%%%%%%%%%%%%
\begin{table}[htbp]
\caption {\small $K^-$-deuteron scattering length (in fm)
calculated using different approximations }

\begin{tabular}{c|cc|c}
approximations & Physical basis    & Isospin basis
&  Isospin basis, Ref.[19]  \\
\hline IA
& $ -0.260  + i\, 1.872 $ & $-0.318 + i\, 2.013 $ & $-0.364 + i\, 1.826 $
\\
\hline IA + double resc.
& $ -2.735  + i\, 2.895 $ & $-3.168 + i\, 3.717 $ & $-2.380 + i\, 1.485 $
\\
\hline IA + double+triple resc.
& $ -3.849  + i\, 2.963 $ & $-5.195 + i\, 4.935 $ & $-2.858 + i\, 0.089 $
\\
\hline $A_{Kd}$ (only el.resc.)
& $  -1.161 + i\, 1.336 $ & $-1.255 + i\, 1.518 $ & $-0.997 + i\, 1.212 $
\\
$A_{Kd}$ (charge exch.)
& $  -0.454 + i\, 0.573 $ & $-0.654 + i\, 0.937 $ & $-0.539 + i\, 0.079 $
\\
$A_{Kd}$ (total)
& $  -1.615 + i\, 1.909 $ & $-1.909 + i\, 2.455 $ & $-1.536 + i\, 1.291 $
\\
\end{tabular}
\label{tab:2}
\end{table}
%%%%%%%%%%%%%%%%%%%%%%%%%%%%%%%%%%%  
%\end{minipage}

\subsection{ $\bar{K}$ in nuclei}
 Next we address the properties of the $\bar{K}$ in the 
nuclear medium which have been studied in  \cite{knuc}. The work is based on
the elementary $\bar{K} N$ interaction which was obtained in \cite{kaon}
using a coupled channel unitary approach with chiral Lagrangians.

The coupled channel formalism requires to evaluate the transition
amplitudes between the different channels that can be built from
the meson and baryon octets. For $K^- p$ scattering there are 10 such
channels, namely $K^-p$, $\bar{K}^0 n$, $\pi^0
\Lambda$, $\pi^0 \Sigma^0$,
$\pi^+ \Sigma^-$, $\pi^- \Sigma^+$, $\eta \Lambda$, $\eta
\Sigma^0$,
$K^+ \Xi^-$ and $K^0 \Xi^0$. In the case of $K^- n$ scattering
the coupled channels are: $K^-n$, $\pi^0\Sigma^-$,
 $\pi^- \Sigma^0$, $\pi^- \Lambda$, $\eta
\Sigma^-$ and
$K^0 \Xi^-$.

 At low energies the transition amplitudes can be written as 
\begin{equation}
V_{i j} = - C_{i j} \frac{1}{4 f^2} (k_j^0 + k_i^0) \ ,
\end{equation}
where $k^{0}_{i,j}$ are the energies of the mesons
and the explicit values of the coefficients $C_{ij}$ can be
found in Ref. \cite{kaon}. 
The coupled-channel Bethe Salpeter equation with the kernel (potential) 
$V_{i j}$
was used in \cite{kaon} in order to obtain the elastic and transition matrix
elements in the $K^- N$ reactions. The diagrammatic expression of this series
can be seen in fig. 6.
\begin{figure}[htb]
 \begin{center}
\includegraphics[height=1.8cm,width=12.cm,angle=0] {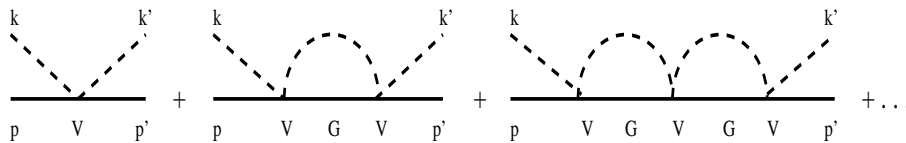}
 \caption{Diagrammatic representation of the Bethe-Salpeter equation.}
 \end{center}
\end{figure}
 The Bethe Salpeter
equations in the center of mass frame read
\begin{equation}
T_{i j} = V_{i j} + {V_{i l} \; G_l \; T_{l j}}  ,
\end{equation}
where the indices $i,l,j$ run over all possible channels and $G_l$ stands
for the loop function of a meson and a baryon propagators.

 In order to evaluate
the $\bar{K}$ selfenergy, one needs first to include the medium modifications 
in the $\bar{K} N$ amplitude, $T_{\rm
eff}^{\alpha}$ ($\alpha={\bar K}p,{\bar K}n$), and then perform the
integral over the nucleons in the Fermi sea: 

\begin{equation}
\Pi^s_{\bar{K}}(q^0,{\vec q},\rho)=2\int \frac{d^3p}{(2\pi)^3}
n(\vec{p}) \left[ T_{\rm eff}^{\bar{K}
p}(P^0,\vec{P},\rho) +
T_{\rm eff}^{\bar{K} n}(P^0,\vec{P},\rho) \right] \ ,
\label{eq:selfka}
\end{equation}

The values
$(q^0,\vec{q}\,)$ stand now for the energy and momentum of the
$\bar{K}$ in the lab frame, $P^0=q^0+\varepsilon_N(\vec{p}\,)$,
$\vec{P}=\vec{q}+\vec{p}$ and $\rho$ is the nuclear matter density.

We also include a p-wave contribution to the ${\bar K}$ 
self-energy coming from the coupling of the ${\bar K}$ meson to
hyperon-nucleon hole ($YN^{-1}$) excitations,
with $Y=\Lambda,\Sigma,\Sigma^*(1385)$. The vertices $MBB^\prime$ are 
easily derived from
the $D$ and $F$ terms of Eq.~(5). The explicit expressions can be seen in
 \cite{knuc}. 
 At this point it is interesting to recall three different approaches to the
 question of the $\bar{K}$ selfenergy in the nuclear medium. The first
 interesting realization was the one in \cite{koch94,wkw96,waas97}, 
 where the Pauli blocking in the intermediate nucleon states in
 fig. 6 induced a shift of the $\Lambda(1405)$ resonance to higher
 energies and a subsequent attractive $\bar{K}$ selfenergy. The work of
 \cite{lutz} introduced a novel an interesting aspect, the selfconsistency.
 Pauli blocking required a higher energy to produce the resonance, but having a
 smaller kaon mass led to an opposite effect, and as a consequence the
 position of the resonance was brought back to the free position. Yet, a 
 moderate
 attraction on the kaons still resulted, but weaker than anticipated from the
 former work.  The work of \cite{knuc} introduces some novelties. It
 incorporates the selfconsistent treatment of the kaons done in \cite{lutz} 
 and in addition it also includes the selfenergy of the pions, which are let to
 excite ph and $\Delta h$ components. It also includes the  mean field
 potentials of the baryons.  
\begin{figure}[htb]
 \begin{center}
\includegraphics[height=8.cm,width=10.cm,angle=0] {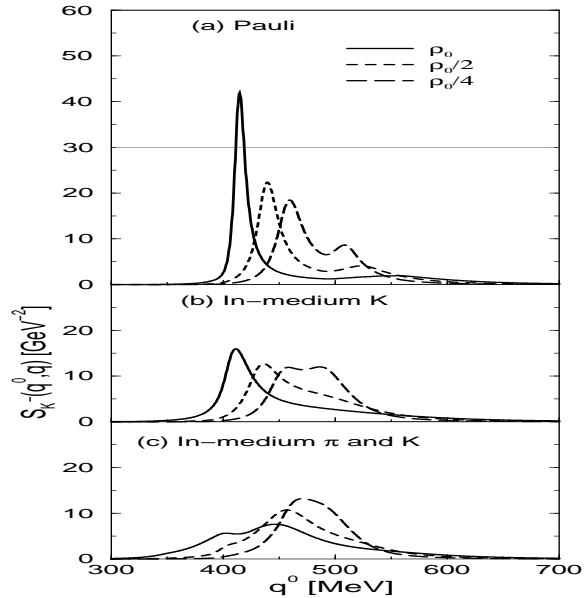}
 \caption{Kaon spectral function at several densities.}
 \end{center}
\end{figure}
The obvious consequence of the work of \cite{knuc}
is that the spectral function of the kaons  gets much wider than in the two
former approaches because one is including new decay channels for the
$\bar{K}$ in nuclei. This can be seen in fig. 7.
The work of \cite{knuc} leads to an attractive potential around nuclear matter
density and for kaons close to threshold of about 40 MeV and a width of 
about 100 MeV.

\section{ Kaonic atoms}

  In the work of \cite{zaki} the kaon selfenergy discussed above has been 
  used for the case of kaonic atoms, where there are
abundant data to test the theoretical predictions. One uses the Klein 
Gordon equation and obtains two families of states. One
family corresponds to the atomic states, some of which are those already  
measured, and 
which have  energies around or below 1 MeV and widths
of about a few hundred KeV or smaller. The other family corresponds to 
states which are nuclear deeply bound states, with
energies of 10 or more MeV and widths around 100 MeV. 
\begin{figure}[htb]
 \begin{center}
\includegraphics[height=7.cm,width=9.cm,angle=0] {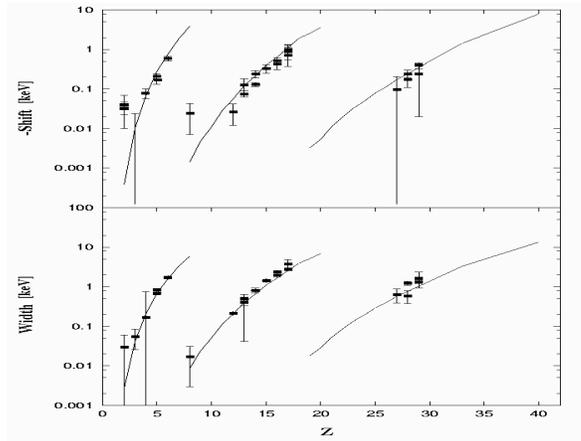}
 \caption{ Shifts and widths of kaonic atoms.}
 \end{center}
\end{figure}
 In fig. 8 we can see the results obtained for shifts and widths for 
 a large set of nuclei around the periodic table. The agreement
with data is sufficiently good to endorse the fairness of the theoretical 
potential. A best fit with a strength of the potential
slightly modified around the theoretical values can lead to even better
 agreement as shown in \cite{baca} and serves to quantify
the level of accuracy of the theoretical potential, which is set there at 
the level of 20-30 per cent as an average. The curious thing
is that there are good fits to the data using potentials with a strength 
at $\rho=\rho_0$ of the order of 200 MeV \cite{gal}. 
As shown in  \cite{zaki}, the results obtained there and those obtained using 
the potential
 of  \cite{gal} are in excellent agreement for
the atomic states. The differences in the two potentials 
appear in the deeply bound nuclear states. The deep potential
provides extra states bound by about 200 MeV, while the potential of 
\cite{knuc} binds states at most by 40 MeV.  This
remarkable finding can be interpreted as saying that the extra bound states, 
forcing the atomic states to be orthogonal to them, introduces
 extra nodes in the wave function and pushes the atomic states more
 to the surface of the nucleus, acting effectively as a
repulsion which counterbalances the extra attraction of the potential.  
This observation also tells that pure fits to the $K^-$
atoms are not sufficient to determine the strength of the $K^-$ nucleus 
potential. Other solutions with even more attraction at 
$\rho=\rho_0$ are in principle possible, provided they introduce new states 
of the deeply bound nuclear family.  On the other
hand, the work of \cite{baca} also tells us that at least an attraction as
 the one provided by the theoretical potential is needed.
 
  The potential used for the $K^-$ atoms has used only the s-wave potential,
ignoring a possible contribution of the p-wave potential and some nonlocalities
associated to the energy and momentum dependence of the s-wave part of the
optical potential. In a recent paper \cite{flor} these corrections have been 
evaluated and
they have been reported as providing a contribution to the potential as large or
even larger than the local s-wave part of the potential used in
\cite{zaki}. In view of the conflict of such a finding with all previous works
on the topic which have systematically relied on the local s-wave potential, we
have reconducted a thorough study of these nonlocalities together with the
contribution of the genuine p-wave part of the potential stemming from the
elementary p-wave amplitude. The work is done in \cite{carmen} and the results
are quite different of those obtained in \cite{flor}, the main reason for the
differences being the inconsistent treatment of the low density limit in
\cite{flor}. The findings of \cite{carmen} lead to corrections from all sources
of nonlocalities which are rather small, smaller than the experimental errors,
and hence this justifies the neglect of such terms in \cite{zaki} and in all
previous studies of kaonic atoms.

\section{ $\phi$ decay in nuclei}

Finally let us say a few words about the $\phi$ decay in nuclei. The work 
reported here \cite{phi} follows closely the lines of
\cite{klingl,norbert}, however, it uses the updated $\bar{K}$ 
selfenergies of \cite{knuc}.  In the present
case the $\phi$ decays primarily in $K\bar{K}$, but these kaons can 
now interact with the medium as discussed previously.    For
the selfenergy of the $K$, since the $KN$ interaction is not too strong 
and there are no resonances, the $t\rho$ approximation is
sufficient.
\begin{figure}[htb]
 \begin{center}
\includegraphics[height=6.cm,width=8.cm,angle=0] {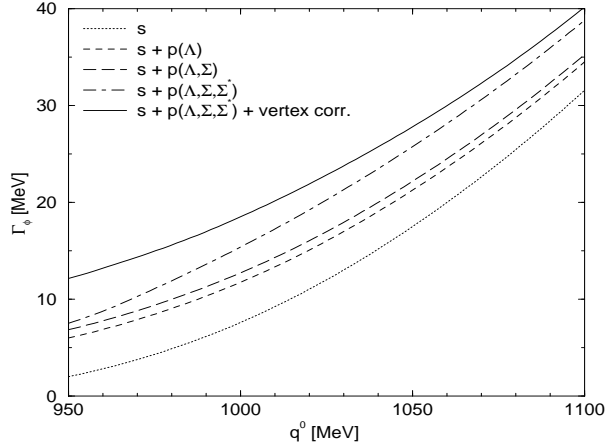}
 \caption{$\phi$ width at $\rho=\rho_0$.}
 \end{center}
\end{figure}
In fig. 9 we show the results for the $\phi$ width at $\rho=\rho_0$ 
as a function of the mass of the $\phi$, separating the
contribution from the different channels. What we observe is that the 
consideration of the s-wave $\bar{K}$-selfenergy is
responsible for a sizeable increase of the width in the medium, but 
the p-wave is also relevant, particularly the $\Lambda h$
excitation and the $\Sigma^*h$ excitation. It is also interesting to 
note that the vertex corrections \cite{beng}  (Yh loops attached
to the $\phi$ decay vertex)  are now present and do not cancel off 
shell contributions like in the case of the scalar mesons. Their
contribution is also shown in the figure and has about the same 
strength as the other p-wave contributions.  The total width of
the $\phi$ that we obtain is about 22 MeV at $\rho=\rho_0$, about 
a factor two smaller than the one obtained in
\cite{klingl,norbert}, yet, the important message is  the 
nearly one order of magnitude increase of the width with respect to
the free one. 
   We are hopeful that in the near future one can measure the width 
   of the $\phi$ in the medium, from
heavy ion reactions or particle nucleus interactions, although it will 
require careful analyses as shown in \cite{indio} for the case
of $K\bar{K}$ production in heavy ion collisions, where consideration 
of the possibility that the observed kaons come from
$\phi$ decay outside the nucleus leads to nuclear $\phi$ widths 
considerably larger than the directly observed ones. 

 In order to facilitate the experimental search of these medium modifications we
 have recently proposed a method based on the $\phi$ photoproduction in nuclei
 \cite{toki} producing $\phi$ with small momenta ( around 150 MeV/c) which
 would be forced to decay inside the nucleus. These slow $\phi$'s would come from
 the elementary $\phi$ photoproduction on a nucleon in the backward direction in
 the CM frame.  A recent measurement of this quantity at Jefferson Lab 
 \cite{jeff} gives
 great hopes that the photoproduction in nuclei with these small $\phi$ momenta 
 could be feasible, since the cross sections obtained, \cite{toki}, are of the 
 same order of magnitude  as those measured in \cite{jeff}. The results of
 \cite{toki} indicate that, once the corrections for the final state interaction 
 of the kaons from the $\phi$ decay are done, one can reconstruct an invariant
 mass of the $\phi$ in the medium at not too high densities but big enough to
 produce a $\phi$ width about twice as big as the free one.

\section{ Summary} 

In summary, we have reported here on recent work which involves the 
propagation of kaons in the nuclear medium. All them
together provide a test of consistency of the theoretical 
ideas and results previously developed and reported here. If we
gain confidence in those theoretical methods one can proceed to 
higher densities and investigate the possibility of kaon
condensates in neutron stars \cite{kaplan}.  The weak strength of our 
$\bar{K}$ potential  would make however the phenomenon highly
unlikely. 
  
On the other hand, we can also extract some conclusions concerning the 
general chiral framework: 
1) The chiral Lagrangians have much information in store. 
2) Chiral perturbation theory allows one to extract some of this 
information. 
3) The chiral unitary approach allows one to extract much more information. 
4) These unitary methods combined with the use of standard many body 
techniques are opening the door to the investigation of
new nuclear problems in a more accurate and systematic way, giving
 rise to a new field which could be rightly called "Chiral
Nuclear Physics". As chiral theory becomes gradually a more accepted
tool to deal with strong interactions at intermediate energies, chiral 
nuclear physics is bound to follow 
analogously in the interpretation of old and new phenomena in
nuclei.

\bigskip

\subsection*{Acknowledgments}
We acknowledge partial financial support from the DGICYT under
contract PB96-0753 and from the EU TMR network Eurodaphne, contract no. 
ERBFMRX-CT98-0169.

\end{document}